\documentclass[10pt,aps]{revtex4}

\usepackage{graphicx}
\usepackage{amsmath} 
\usepackage{textcomp}
\usepackage{amsfonts}

\everymath{\displaystyle}

\newcommand{\barr}{\begin{eqnarray}}
\newcommand{\earr}{\end{eqnarray}}

\newcommand{\Etilde}[2]{A_{#1}(\mathbf{#2})}

\newcommand{\EtildeC}[2]{A^*_{#1}(\mathbf{#2})}

\newcommand{\campo}[1]{\mathbf{#1}(\mathbf{r},t)}
\newcommand{\vett}[1]{\mathbf{#1}}
\newcommand{\uvett}[1]{\hat{\vett{#1}}}
\newcommand{\bra}[1]{\langle #1|}
\newcommand{\ket}[1]{|#1\rangle}
\newcommand{\baseE}[2]{\uvett{e}_{#1}(\vett{#2})}

\newcommand{\baseEC}[2]{\uvett{e}_{#1}^*(\vett{#2})}
\newcommand{\GH}{Goos-H$\mathrm{\ddot{a}}$nchen }

\begin{document}
\title{Goos-H$\mathbf{\ddot{a}}$nchen and Imbert-Fedorov shifts for bounded wave packets of light}

\author{Marco Ornigotti}
\affiliation{Max Planck Institute for the Science of Light, G$\ddot{\mathrm{u}}$nther-Scharowsky-Stra$\ss$e 1/Bau24, 91058 Erlangen, Germany} 
\email{marco.ornigotti@mpl.mpg.de}

\author{Andrea Aiello}
\affiliation{$^1$Max Planck Institute for the Science of Light, G$\ddot{\mathrm{u}}$nther-Scharowsky-Stra$\ss$e 1/Bau24, 91058 Erlangen, Germany} 
\affiliation{$^2$Institute for Optics, Information and Photonics, University of Erlangen-N$\ddot{u}$rnberg, Staudtstra$\ss$e 7/B2, 91058 Erlangen, Germany}

\begin{abstract}
We present precise expressions of the spatial and angular \GH and Imbert-Fedorov shifts experienced by a longitudinally and transversally limited beam of light (wave packet) upon reflection from a dielectric interface, as opposed to the well-known case of a monochromatic beam which is bounded in transverse directions but infinitely extended along the direction of propagation. This is done under the assumption that the detector time is longer than the temporal length of the wave packet (wave packet regime). Our results will be applied to the case of a Gaussian wave packet and show that, at the leading order in the Taylor expansion of reflected-field amplitudes, the results are the same of the monochromatic case.
\end{abstract}

\maketitle

\section{Introduction}
The problem of the reflection of an electromagnetic field upon a planar surface is an old and well-known problem of optics which is usually treated by means of the so-called Fresnel reflection (and transmission) coefficients. These  are typically given for the simplest case of impinging plane waves solely \cite{mandel, jackson, born}. For beam-like field configurations this is not a problem since any electromagnetic field of arbitrary shape can be decomposed into plane waves. Thus, the reflected field can be evaluated by first reflecting each of its plane wave components individually and then sum them back. However, non-specular reflection effects may occur and be observed \cite{ref4,ref6,ref7}. Among all these effects, the most famous ones are the so-called Goos-H$\mathrm{\ddot{a}}$nchen (GH) \cite{ref8GH,ref9GH,ref8,ref9,ref10} and Imbert-Fedorov (IF) \cite{ref12IF,ref13IF,ref12,ref13} shifts, the former occurring in the plane of incidence of light while the latter occurs in the plane orthogonal to the plane of incidence. Detailed theoretical predictions and experimental observation of both GH and IF shifts have been given in the past decades for different types of surfaces (dielectric, metallic \cite{ref11,refA}, multilayered \cite{ref3,ref5}, et cetera) and for different beam shapes (Hermite-Gauss \cite{ref2,refD}, Laguerre-Gauss \cite{refC}, Bessel \cite{refB} et cetera). Despite the vast amount of literature on the subject, however, there is almost  no clear and simple theory for the non-monochromatic case, namely for beams of light which have not only finite transverse size, but also finite temporal extension (wave packets). A notable exception is represented by the work of Gragg \cite{ref7} whose treatment is however limited to the case of scalar waves.

In this work, starting from the work done in Ref. \cite{aielloW} and following the formalism of beam shifts as displacements of the center of mass of the energy density distribution of the electromagnetic field upon reflection introduced in Ref. \cite{andreaOL,andreaNJP}, we extend the concepts of \GH and Imbert-Fedorov shifts to the non-monochromatic case, firstly by generalizing the usual monochromatic formulas to the non-monochromatic case, and then by applying these new results to the simple case of a Gaussian quasi monochromatic wave packet.

This work is organized as follows: in Sect. 2 the notation used throughout the manuscript is established and the plane-wave expansion of a non-monochromatic field is given. This  will be used in Sect. 3 to calculate the center of mass of the field energy density distribution, which will allow us to calculate the shifts. Then, in Sect. 4 a brief recall on how to calculate the reflected field amplitudes is given, and the general expression for the non monochromatic center of mass is derived explicitly. In Sect. 5, the application of this formalism to the simple case of a quasi monochromatic Gaussian wave packet is discussed. Finally, in Sect. 6 conclusions are drawn.

\section{Fourier representation of a non-monochromatic field}
\subsection{Notation}
Throughout the manuscript, the following conventions are understood. According to Ref. \cite{berry}, we indicate with capital letters the transversal components of a three dimensional vector, so that we will write $\mathbf{r}=\{x,y,z\}=\{\mathbf{R},z\}$, being $\mathbf{R}=\{x,y\}$ for the position vector and $\mathbf{k}=\{k_x,k_y,k_z\}=\{\mathbf{K},k_z\}$, being $\mathbf{K}=\{k_x,k_y\}$ for the $k$-vector. Moreover, we define $r=|\vett{r}|$, $R=|\vett{R}|$, $k=|\vett{k}|$ and $K=|\vett{K}|$. We also define the gradient operator in the $k$-space as follows:
\begin{equation}
\frac{\partial}{\partial\mathbf{k}}=  \left\{\frac{\partial}{\partial k_x},\frac{\partial}{\partial k_y},\frac{\partial}{\partial k_z}\right\}\equiv\left\{\frac{\partial}{\partial\mathbf{K}},\frac{\partial}{\partial k_z}\right\}.
\end{equation}

Finally we assume the following condition on the summation indexes:  Greek indexes can assume only two values, namely $\mu,\nu, \tau, \cdots \in\{1,2\}$ while Latin indexes assume three values: $i,j,l,\cdots \in\{1,2,3\}$.

\subsection{Local basis}
Let us consider a Cartesian reference frame $(Oxyz)$ associated to the orthonormal basis $\{\uvett{x},\uvett{y},\uvett{z}\}$. Let $\vett{k}_0=k_0\uvett{z}$ be the central wave vector of a wave packet propagating along  the $z$-axis, and $\vett{k}=k_x\uvett{x}+k_y\uvett{y}+k_z\uvett{z}$ denotes the generic wave vector with $k=\sqrt{k_x^2+k_y^2+k_z^2}\neq k_0$. Given $\uvett{k} = \vett{k}/k$ and the unit vector $\uvett{z}$ parallel to the $z$-axis,  it is possible to build a complete orthogonal basis $\uvett{e}_j(\vett{k})$ out of the standard Cartesian basis by means of a rotation around the axis
\begin{equation}
\uvett{n}=\frac{\uvett{z}\times\vett{k}}{|\uvett{z}\times\vett{k}|},
\end{equation}
by the angle $\theta=\arcsin(K/k)$ between $\vett{k}$ and $\uvett{z}$ \cite{marquardt}, where $``\times''$ denotes the ordinary cross-product in $\mathbb{R}^3$. By representing such a rotation with the Rodrigues' formula \cite{rodrigues}
\begin{equation}
R(\theta,\uvett{n})=e^{\theta M}=I+M\sin\theta+M^2(1-\cos\theta),
\end{equation}
where $I$ is the (three dimensional) identity matrix and the matrix elements of $M$ are given by:
\begin{equation}
M_{ij}=-\sum_{k=1}^3\varepsilon_{ijk}(\uvett{n})_k,
\end{equation}
where $\varepsilon_{ijk}$ is the completely antisymmetric Levi-Civita symbol. By doing this operation the local basis $\uvett{e}_i(\vett{k})$ can be explicitly written as $\baseE{1}{k}=R(\theta,\uvett{n})\uvett{x}$, $\baseE{2}{k}=R(\theta,\uvett{n})\uvett{y}$ and $\baseE{3}{k}=R(\theta,\uvett{n})\uvett{z}\equiv\uvett{k}$.
It is worth noticing that this basis constitutes a local basis attached to the field itself, and that $\baseE{3}{k}$ points in the direction of the wave vector $\vett{k}$, while the other two basis vectors $\baseE{1,2}{k}$ are orthogonal to $\vett{k}$ and they are natural candidates for describing the field polarization \cite{mandel}. These basis vectors have the following properties:
\begin{subequations}
\begin{align}
\baseE{\mu}{k}\cdot\baseE{\nu}{k}= & \;\delta_{\mu\nu}, \\
\baseE{i}{k}\times\baseE{j}{k}= & \;\varepsilon_{ijl}\baseE{l}{k}, \\
\left(\uvett{k}\times\baseE{\mu}{k}\right)\cdot\left(\uvett{k}\times\baseE{\nu}{k}\right)= & \;\delta_{\mu\nu},
\end{align}
\end{subequations}
where the dot symbol indicates the ordinary scalar product in $\mathbb{R}^3$ and summation over repeated indexes is understood. Note that with a suitable unitary transformation \cite{mandel} it is possible to relate this basis with other basis commonly used in optics, like the helicity basis.

\subsection{3D Fourier representation of the electromagnetic field}
Having established the form of the local basis attached to the (incident) beam field, we can now write the electric and magnetic fields in terms of plane waves as follows \cite{mandel}:
\begin{subequations}\label{EandB}
\begin{align}
\vett{E}(\vett{r},t)=& \;\sum_{\mu=1}^2\int d^3k \Etilde{\mu}{k}\baseE{\mu}{k}e^{i(\vett{k}\cdot\vett{r}-\omega t)} + \text{c.c},\label{fieldE} \\
c \, \vett{B}(\vett{r},t)=& \;\sum_{\mu=1}^2\int d^3k \Etilde{\mu}{k}\big[\uvett{k}\times\baseE{\mu}{k}\big]e^{i(\vett{k}\cdot\vett{r}-\omega t)} + \text{c.c}, \label{fieldB}
\end{align}
\end{subequations}
where $\omega=c|\vett{k}|=c\sqrt{k_x^2+k_y^2+k_z^2}$, $d^3k=dk_xdk_ydk_z$ and ``c.c." stands for complex conjugate. The complex amplitude $\Etilde{\mu}{k}$ contains informations on both the envelope and on the polarization of the single spectral component of the field, and can therefore be written as $\Etilde{\mu}{k}\equiv A(\vett{k},\vett{k}_0)\alpha_{\mu}(\vett{k})$, where $A(\vett{k},\vett{k}_0)$ is the scalar distribution of the plane waves components of the wave packet around the central wave vector $\vett{k}_0$ and $\alpha_{\mu}(\vett{k})$ contains informations about the field polarization. 

As a final remark to this section, it is worth noticing that in the integrals that appear in Eqs. \eqref{EandB}, $k_x$, $k_y$ and $k_z$ are \emph{independent} variables. In this general formulation, no hypothesis on the nature of the field (whenever the field is homogeneous or contains evanescent contributions) is made, and consequently no relation between the various components of $\vett{k}$ has been established yet. 

\section{Center of mass of the field energy density}
In this section we will generalize the formalism introduced in Ref. \cite{andreaNJP} for the calculation of monochromatic beam shifts. The essential ingredient is the center of mass of the energy density distribution, defined as follows:
\begin{equation}\label{cdm}
\langle\vett{R}\rangle=\displaystyle\frac{\displaystyle{\int_{-T}^{T}dt \displaystyle\int \,d^2R\,\vett{R}\,\mathcal{H}(\vett{R},z,t)}}{\displaystyle{\int_{-T}^{T}dt\displaystyle\int d^2R\,\mathcal{H}(\vett{R},z,t)}},
\end{equation}
where $\mathcal{H}=\varepsilon_0(\vett{E}^2+c^2\vett{B}^2)/2$ is the electromagnetic field energy density and $T$ is the integration time of the detector. Note that, compared with the expression for a monochromatic field \cite{altroAiello}, here the presence of the integral on time is imposed by the non monochromatic nature of the field. In order for the detector to be able to detect the whole energy of the wave packet (WP), the integration time $T$ must be chosen in such a way to be larger than the length in time of the WP. If we call $\zeta$ the width of the WP along the $z$ direction, with $\zeta k_0\ll 1$ \cite{blow}, $T$ must be compared with $\zeta/c$. Thus, if the integration time is much longer than the temporal length of the WP (i.e., $cT\gg\zeta$), we are permitted to let $T\rightarrow\infty$ in the integrals in Eqs. \eqref{cdm}. Conversely, if the integration time is comparable with the WP timescale (namely if $cT\leq\zeta$), we are not allowed to extend the integration extrema of the time integral to infinity, as the detector can detect only a fraction of the energy of the WP. In this case, then, the role of the detector will be actively important, since the choice of the integration time can drastically modify the properties of the detected field. In this work we will focus our attention on the first case (the so-called \emph{wave packet} regime), leaving the other case (the \emph{beam} regime) to future analysis.

With this in mind let us derive the expression for the center of mass $\langle\vett{R}\rangle$. The electromagnetic field energy density is given by \cite{jackson}
\begin{equation}\label{energia}
\mathcal{H}\equiv\mathcal{H}_e+\mathcal{H}_b=\frac{\varepsilon_0}{2} \left[\campo{E}\cdot\campo{E}+ c^2 \campo{B}\cdot\campo{B} \right].
\end{equation}
By substituting the expressions of the electric and magnetic fields in terms of their spectral component as defined in Eqs. \eqref{EandB} we obtain, after a long but straightforward calculation, the following expression for the center of mass:
\begin{equation}\label{centro}
\langle\vett{R}\rangle=\frac{\displaystyle{\sum_{\mu,\nu}^{1,2}\int \! d^3k \frac{k}{k_z} \EtildeC{\mu}{k} \, \baseEC{\mu}{k} \! \cdot \! 
\left\{-\frac{1}{i}\left[\frac{\partial}{\partial\vett{K}}-\frac{\vett{K}}{k_z}\left(\frac{\partial}{\partial k_z} - \frac{1}{k_z}\right)\right] \! +z\frac{\vett{K}}{k_z} \right\}
 \! \baseE{\nu}{k}\Etilde{\nu}{k}}}{\displaystyle{\sum_{\mu=1}^2\int d^3 k \frac{k}{k_z} \left| \Etilde{\mu}{k} \right|^2}},
\end{equation}
where $k/k_z\geq 0$ plays the role of a weight function that approaches the value $1$ for well collimated beams such that $(k_x^2+k_y^2)/k_z^2\ll 1$.  In obtaining this expression, some care has to be put when integrating over the transverse coordinates and over time. While the integration over $d^2R$ produces a Dirac delta on the transverse $k$-vector $\vett{K}$, leaving an explicit dependence on $k_z$, namely
\begin{equation}\label{relazione}
\int\,\int d^2R\, e^{i(\vett{k}-\vett{k'})\cdot\vett{r}}=(2\pi)^2e^{i (k_z-k_z') z}\delta(\vett{K'}-\vett{K}),
\end{equation}
the time integration has to be analyzed more carefully. To this aim, we will now show how to pass from Eq. \eqref{cdm} to Eq. \eqref{centro}. To do that, we will concentrate on the contribution of the electric part of the electromagnetic energy density to the center of mass \eqref{cdm}, as the magnetic contribution can be obtained using similar calculations. Let us start by evaluating the denominator of Eq. \eqref{cdm}. By substituting the expression of the electric field given by  Eq. \eqref{fieldE} into the expression of the electric part of the electromagnetic energy density that appears in the denominator of Eq. \eqref{centro} we have:
\barr
I_{e}&\equiv &
\int dt \, \iint d^2R\,\mathcal{H}_e(\vett{R},z,t)  \nonumber\\
&=&\frac{\varepsilon_0}{2(2\pi)^3}\sum_{\lambda,\mu}\int d^3k\int d^3k' \EtildeC{\mu}{k}\Etilde{\lambda}{k'}\baseEC{\mu}{k}\cdot\baseE{\lambda}{k'}\nonumber\\
&\times&\int dt e^{-i(\omega-\omega')t}\int d^2R e^{i(\vett{k}-\vett{k'})\cdot\vett{r}}+\text{c.c.},
\earr
where the ``fast-oscillating'' terms leading to $\delta (\omega+\omega') \rightarrow 0$ have been omitted.
This expression can be evaluated by using Eq. \eqref{relazione} and  the standard properties of the Dirac delta function \cite{deltafx} yielding to:
\begin{equation}
\int dt e^{-i(\omega-\omega')t}= 2\pi\delta(\omega'-\omega)=\left(\frac{2\pi}{c}\right)\frac{k}{|k_z|}\Big[\delta(k_z'-k_z)+\delta(k_z'+k_z)\Big],
\end{equation}
where the two Dirac delta functions $\delta(k_z'\pm k_z)$ correspond respectively to the forward and the backward propagating solution. Without any loss of generality, we assume to consider only forward propagating waves (for that we can write $|k_z|=k_z$).
Substituting these expressions into the previous equation leads to the following result:
\barr
I_e&=&\frac{\varepsilon_0}{c}\sum_{\lambda,\mu}\int d^3k\int d^3k'\frac{k}{k_z}\EtildeC{\mu}{k}\Etilde{\lambda}{k'}\baseEC{\mu}{k}\cdot\baseE{\lambda}{k'}\nonumber\\
&\times&e^{i(k_z-k_z')}\delta(\vett{K}'-\vett{K})\delta(k_z'-k_z)\equiv\frac{\varepsilon_0}{c}\sum_{\mu=1}^2\int d^3 k \frac{k}{k_z} \left| \Etilde{\mu}{k} \right|^2.
\earr
For the denominator, then, the presence of the time integration is not critical, since the only effect is to introduce a Dirac delta in the field frequencies $\delta(\omega'-\omega)$ that 
is easily transformed into a Dirac delta in the $k_z$ component of the $k$-vector times the factor $k/k_z$.

For the numerator, the calculations need a bit more of attention. First of all, for the sake of simplicity, let us consider only the contribution given by $x\campo{E}$, since the contribution of the term $y\campo{E}$ together with its magnetic counterpart can be straightforwardly calculated with similar techniques. By calling $I_{ex}\equiv\int dt\int\int d^2R\, x\,\mathcal{H}_e(\vett{R},z,t)$ and by remembering that (when integrating in $d^3k$) the presence of the multiplicative term $x$ can be turned into a derivative with respect to $k_x$ by means of part integration we obtain the following result:
\barr\label{starting_point}
I_{ex}&=&\frac{\varepsilon_0}{2(2\pi)^3}\sum_{\lambda,\mu}\int d^3k\int d^3k'\int dt\int\int d^2R\,\EtildeC{\mu}{k'}\baseEC{\mu}{k'}e^{i(\vett{k'}\cdot\vett{r}-\omega't)}\nonumber\\
& &\times \left\{- \frac{1}{i}\frac{\partial}{\partial k_x}\Bigg[\Etilde{\lambda}{k}\baseE{\mu}{k}e^{i(\vett{k}\cdot\vett{r}-\omega t)}\Bigg]+\text{c.c}\right\},
\earr
where, again, the ``fast oscillating'' terms have been omitted as they give zero contribution.
Note that the calculation of the derivative brings four terms. While the first three of them (the derivative of the amplitude, the derivative of the local basis and the derivative of the spatial plane wave respectively) have a time dependence of the form $\exp{[-i(\omega'-\omega)t]}$ that leads, once integrated, to a Dirac delta in the field frequencies $\delta(\omega'-\omega)$, the fourth term contains the derivative of $\omega$ with respect to $k_x$ (since $\omega=c|\vett{k}|$). This brings down a factor of $t$ from the time exponential that gives rise (once integrated in time) to a derivative of the Dirac delta with respect to $\omega$, i.e., $\partial\delta(\omega'-\omega)/\partial\omega$. 

Since $\omega$ is not an integration variable but it depends on $|\vett{k}|$ via the vacuum dispersion relation $\omega=c\sqrt{k_x^2+k_y^2+k_z^2}\equiv c k$ (and equivalently $\omega'=c\sqrt{k_x^2+k_y^2+k_z'^2}\equiv ck'$),  it is more convenient to express this Dirac delta as a function of the variable $k_z$, in order to relate it to one of the integration variables and use it lately to eliminate the integration in $k_z'$. By using Eq. \eqref{relazione} together with the following formula \cite{deltafx} 
\begin{equation}
\int dt \, t \,  e^{-i(\omega-\omega')t}= -\left(\frac{2\pi}{ic}\right)\frac{kk'}{k k_z'+k' k_z}\left(\frac{\partial}{\partial k_z'}-\frac{\partial}{\partial k_z}\right)\delta(\omega'-\omega),
\end{equation}
it is possible, after some straightforward algebra, to obtain the following result:
\barr
I_{ex}&=&\frac{1}{c}\sum_{\lambda,\mu}\int d^3k\frac{k}{k_z}\EtildeC{\lambda}{k}\Bigg\{\delta_{\lambda\mu}\Bigg[-\frac{1}{i}\Bigg(\frac{\partial}{\partial k_x}-\frac{k_x}{k_z}\frac{\partial}{\partial k_z} + \frac{k_x}{k_z^2}\Bigg)+z\frac{k_x}{k_z}\Bigg]\nonumber\\
&+&\baseEC{\lambda}{k}\cdot\Bigg[-\frac{1}{i}\Bigg(\frac{\partial}{\partial k_x}-\frac{k_x}{k_z}\frac{\partial}{\partial k_z}\Bigg)\baseE{\mu}{k}\Bigg]\Bigg\}\Etilde{\mu}{k}.
\earr
The same result can be obtained for the $y$-component by making the substitution $k_x\rightarrow k_y$.

In order to make Eq. \eqref{centro} more easy to understand and manage, it is possible to exploit the quantum notation for wave packet and beams of light introduced in Ref. \cite{alonso} to rewrite Eq. \eqref{centro} in a more appealing way by separating the contributions to the center of mass of the energy distribution in three parts as follows:
\begin{equation}\label{corta}
\langle\vett{R}\rangle=\langle\vett{R}\rangle_S+\langle\vett{R}\rangle_B+z\langle\vett{R}\rangle_A,
\end{equation}
where the first term accounts for spatial shifts, the second term for the contribution of the Berry connection \cite{alonso} and the third $z$-dependent term accounts for the angular shift. Their explicit form is given by:
\begin{subequations}\label{parti}
\begin{align}
\langle\vett{R}\rangle_S= & \; \frac{1}{N}\sum_{\mu}\bra{A_{\mu}}\frac{k}{k_z}\left\{ -\frac{1}{i}\left[\frac{\partial}{\partial\vett{K}}-\frac{\vett{K}}{k_z}\left( \frac{\partial}{\partial k_z} - \frac{1}{k_z} \right)\right] \right\}\ket{A_{\mu}},
\label{RS}\\
\langle\vett{R}\rangle_B=& \;\frac{1}{N}\sum_{\mu,\nu}\bra{A_{\mu}}\frac{k}{k_z}\Lambda_{\mu\nu}(\vett{k})\ket{A_{\nu}},
\label{RB} \\
\langle\vett{R}\rangle_A=& \;\frac{1}{N}\sum_{\mu}\bra{A_{\mu}}\frac{k}{k_z}\left(\frac{\vett{K}}{k_z}\right)\ket{A_{\mu}}, \label{RA}
\end{align}
\end{subequations}
where
\begin{equation}
\Lambda_{\mu\nu}(\vett{k})=\baseEC{\mu}{k}\cdot\left[-\frac{1}{i} \left(\frac{\partial}{\partial\vett{K}}-\frac{\vett{K}}{k_z}\frac{\partial}{\partial k_z} \right)\right]\baseE{\nu}{k}
\end{equation}
is the Berry connection expressed in the basis $\baseE{\mu}{k}$ and
\begin{equation}
N=\sum_{\mu}\bra{A_{\mu}}\frac{k}{k_z}\ket{A_{\mu}}
\end{equation}
is the normalization factor (i.e., the denominator of Eq. \eqref{centro}). This is the first main result of this work. As can be seen from the previous equations, a first principal difference with respect to the monochromatic case is the presence of the weighting factor $k/k_z$. This term, that comes from the conversion of the Dirac delta  $\delta(\omega'-\omega)$ (yield by time integration) into a Dirac delta $\delta(k_z'-k_z)$, takes into account the fact that the component $k_z$ of the wave vector $k$ can now independently take any value and it is not anymore constrained to be dependent on $k_x$ and $k_y$ as in the monochromatic case.

The other important difference with respect to the monochromatic case is the presence of the last factor (the term $1/k_z$ in the round parenthesis of Eq. \eqref{RS}). In order to understand the origin of this term, let us consider explicitly the calculation of  Eq. \eqref{RS}. 

Starting from Eq. \eqref{starting_point}, after having performed the integration in the transverse spatial coordinates and the two time integrations (one that gives $\delta(\omega'-\omega)$ and the other one that gives $\partial\delta(\omega'-\omega)/\partial\omega'$) we obtain the following expression (we drop the constant factors and the summation over $\mu$ since they are of no relevance for this discussion):
\barr
\displaystyle{\int d^3k \frac{k}{|k_z|}\Bigg\{\Bigg[\EtildeC{\mu}{k}\left(-\frac{1}{i}\frac{\partial}{\partial\vett{K}}\right)\Etilde{\mu}{k}+\Etilde{\mu}{k}\left(\frac{1}{i}\frac{\partial}{\partial\vett{K}}\right)\EtildeC{\mu}{k}\Bigg]}\nonumber\\
+\frac{k_x}{k_z}\Bigg[A_{\mu}\Bigg(\frac{1}{i}\frac{\partial}{\partial k_z}\Bigg)A^*_{\mu}+A_{\mu}^*\Bigg(-\frac{1}{i}\frac{\partial}{\partial k_z}\Bigg)A_{\mu}\Bigg]\Bigg\}
\earr
where the $z$-dependent part and the term that contains the Berry connection have been ignored since the last term only originates from these two terms.
By noting that (since $|\Etilde{\mu}{k}|^2$ is real)
\begin{equation}
\Etilde{\mu}{k}\left(\frac{1}{i}\frac{\partial}{\partial k_j}\right)\EtildeC{\mu}{k}=\frac{1}{i}\frac{\partial}{\partial k_j}|\Etilde{\mu}{k}|^2-\EtildeC{\mu}{k}\left(\frac{1}{i}\frac{\partial}{\partial k_j}\right)\Etilde{\mu}{k},
\end{equation}
where $k_j\in\{k_x,k_y,k_z\}$, the previous term can be rewritten as follows:
\begin{equation}
2\int d^3k \frac{k}{|k_z|}\EtildeC{\mu}{k}\Bigg[-\frac{1}{i}\left(\frac{\partial}{\partial\vett{K}}-\frac{\vett{K}}{k_z}\frac{\partial}{\partial k_z}\right)\Bigg]\Etilde{\mu}{k}+\int d^3k \frac{k}{|k_z|} \Bigg[\frac{1}{i}\Bigg(\frac{\partial}{\partial\vett{K}}-\frac{\vett{K}}{k_z}\frac{\partial}{\partial k_z}\Bigg)\Bigg]|\Etilde{\mu}{k}|^2.
\end{equation}
The first integral gives the usual term that depends on the linear momentum, while the second term, after a partial integration, accounts exactly for the extra term in Eq. \eqref{RS}. This extra term, then, is due to the non monochromatic nature of the field, since it involves, ultimately, the derivative of the weighting factor ${k}/{|k_z|}$, namely
\begin{equation}
-\frac{1}{i}\Bigg(\frac{\partial}{\partial\vett{K}}-\frac{\vett{K}}{k_z}\frac{\partial}{\partial k_z}\Bigg)\frac{k}{|k_z|}.
\end{equation}
 In the monochromatic case, in fact, since the weighting factor is constant, this term vanishes.

\section{Calculation of the reflected amplitudes}
\begin{figure}[t!]
\begin{center}
\includegraphics[width=0.8\textwidth]{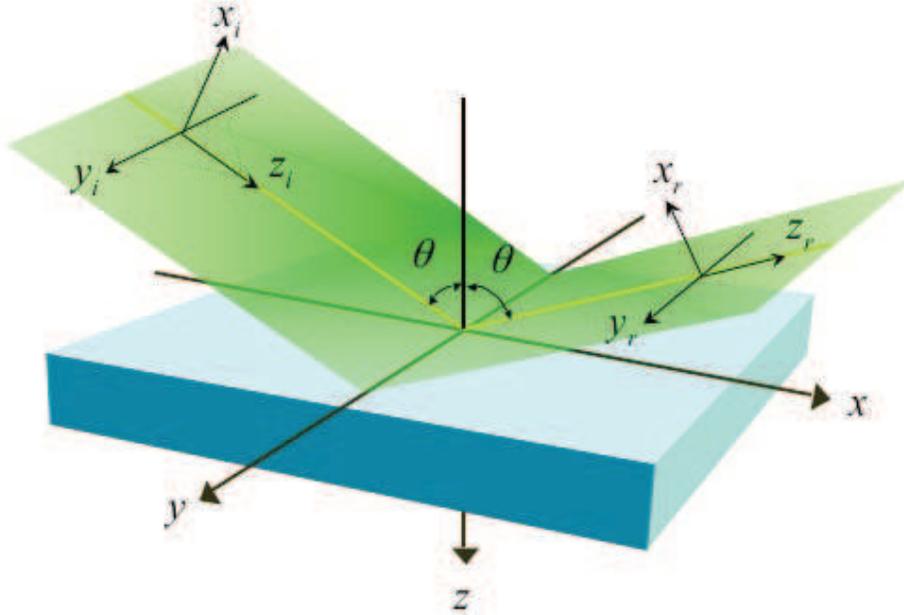}
\caption{Geometry of the problem. $\{\uvett{x}_i,\uvett{y}_i,\uvett{z}_i\}$ is the reference frame attached to the incident wave packet, $\{\uvett{x}_r,\uvett{y}_r,\uvett{z}_r\}$ is the reference frame attached to the reflected wave packet and $\{\uvett{x},\uvett{y},\uvett{z}\}$ is the laboratory frame. $\theta$ is the angle of incidence.}
\label{riflessione}
\end{center}
\end{figure}

The geometry of the problem is depicted in Fig. \ref{riflessione}. We indicate with $\vett{k}$ the representation of the $k$-vector in the reference frame attached with the incident field, while with $\tilde{\vett{k}}$ we intend the representation of the $k$-vector in the frame attached to the reflected beam \cite{ref7}.  Upon reflection, the relation between these two $k$-vectors is the following:
\begin{equation}
 \tilde{\vett{k}}=k_x\uvett{x}+k_y\uvett{y}-k_z\uvett{z}.
 \end{equation}
 The following rules for scalar product also trivially hold: $\tilde{\vett{k}}\cdot\vett{r}=\vett{k}\cdot\tilde{\vett{r}}$ and $\vett{k}\cdot\vett{r}=\tilde{\vett{k}}\cdot\tilde{\vett{r}}$. Having written the electric and magnetic fields in terms of their 3D Fourier spectral components, all that is needed for calculating the amplitudes of the reflected field is to know how the single plane wave components reflect. Following Ref. \cite{aielloW} it is easy to show that in order to calculate the reflected field amplitude it is enough to do the following substitution:
\begin{equation}\label{reflex}
\Etilde{\mu}{k}\baseE{\mu}{k}e^{i\vett{k}\cdot\vett{r}}\rightarrow \Etilde{\mu}{k}r_{\mu}(\vett{k})\baseE{\mu}{\tilde{k}}e^{i\vett{k}\cdot\tilde{\vett{r}}} \equiv A_{\mu}'(\vett{k})\baseE{\mu}{\tilde{k}}e^{i\vett{k}\cdot\tilde{\vett{r}}},
\end{equation}
where $r_{\mu}(\vett{k})$ is the Fresnel reflection coefficient for the $p$-polarized waves (or TM polarization, that is plane waves with the electric field parallel to the plane of incidence) for $\mu=1$ and for $s$-polarized waves (or TE, that is plane waves with the electric field orthogonal to the plane of incidence) for $\mu=2$. The explicit expression for the Fresnel reflection coefficient is given by \cite{mandel,aielloW}
\begin{subequations}
\begin{align}
r_p(\vett{k})=& \; \frac{\epsilon k_z-k_z^t}{\epsilon k_z+k_z^t},\nonumber\\
r_s(\vett{k})=& \; \frac{k_z-k_z^t}{k_z+k_z^t},
\end{align}
\end{subequations}
where $\epsilon$ is the (generally complex-valued) relative dielectric constant $\epsilon_1/\epsilon_2$ of the two media that constitute the interface and $k_z^t=\sqrt{\epsilon k-k_x^2-k_y^2}$ is the $z$-component of the wave vector inside the medium. Note that these expressions are conceptually different from the one normally given in textbooks \cite{mandel, born} for the monochromatic case. Here, the term $k$ is no longer constant and it cannot be factorized outside the square root anymore. Thus, in the expression of the reflection coefficients for a non monochromatic fields, an intrinsic dependence of the coefficient on the frequency $\omega$ (or analogously on the $k_z$ component of the wave vector) is present.

The expression for the center of mass of the energy distribution in the reflected frame is then equal to the one obtained in the incident frame (Eq. \eqref{centro}) with the only difference that now we have to consider as weighting functions the amplitudes of the field in the reflected frame, namely we need to make the substitution $\ket{A_{\mu}} \rightarrow \ket{A_{\mu}'}$ and $\bra{A_{\mu}}\rightarrow\bra{A_{\mu}'}$. Then, following the prescription given in Ref. \cite{andreaNJP}, given a center of mass distribution $\langle \vett{R}\rangle(z)$ like the one calculated in Eq. \eqref{centro} it is possible to compute both the spatial and angular \GH and Imbert-Fedorov shifts by means of the following relation:
\begin{subequations}\label{formula}
\begin{align}
 \Delta_\text{GH}^{\text{sp}} \equiv \Big. \langle x \rangle \Big|_{z=0}, \qquad \Delta_\text{GH}^{\text{ang}} \equiv \frac{\partial}{\partial z}\langle x \rangle, \\
 \Delta_\text{IF}^{\text{sp}} \equiv \Big. \langle y \rangle \Big|_{z=0}, \qquad \Delta_\text{IF}^{\text{ang}} \equiv \frac{\partial}{\partial z}\langle y \rangle.
\end{align}
\end{subequations}
Note that in our case, by using the expression given by Eq. \eqref{corta} for the center of mass distribution and this definition of the two shifts, the expression of the center of mass can be put into a direct relation with the \GH and Imbert-Fedorov shifts thanks to the following relation:
\begin{equation}
\langle\vett{R}\rangle=\uvett{x} \, \Delta_\text{GH}(z)+\uvett{y} \, \Delta_\text{IF}(z),
\end{equation}
where, according to Eqs. \eqref{formula}
\begin{subequations}
\begin{align}
\Delta_{\text{GH}}(z)=\langle x \rangle_S+\langle x \rangle_B+z\langle x\rangle_A\equiv \Delta_{\text{GH}}^{\text{sp}}+z\Delta_{\text{GH}}^{\text{ang}},\\
\Delta_{\text{IF}}(z)=\langle y\rangle_S+\langle y\rangle_B+z\langle y\rangle_A\equiv \Delta_{\text{IF}}^{\text{sp}}+z\Delta_{\text{IF}}^{\text{ang}}.
\end{align}
\end{subequations}
Note that this expression that connects the center of mass with the shifts is very general (namely, it is beam-shape independent). The explicit expression of both \GH and Imbert-Fedorov shifts can be calculated once the explicit shape of the beam is known. 

\section{Application to a Gaussian wavepacket}
In this section we will now apply the formula given above for the case of Gaussian wave packet. We assume that it is possible to write this wave packet as a narrow distribution of $k$-vectors around a central value $\vett{k}_0$ (quasi monochromatic approximation). We also define in the reference frame $(Ox_iy_iz_i)$ attached to the incident beam the vector $\uvett{f}=a_p\uvett{x}_i+a_s\exp{(i\phi)}\uvett{y}_i$  that accounts for the beam polarization, with $a_p, a_s, \phi \in \mathbb{R}$ and $a_p^2+a_s^2=1$. The expression of the polarization function $\alpha_{\mu}(\vett{k})$ is then given by
\begin{equation}
\alpha_{\mu}(\vett{k})=\Bigg[1-\frac{|\uvett{f}\cdot\vett{k}|^2}{k^2}\Bigg]^{-1/2}\baseE{\mu}{k}\cdot\uvett{f},
\end{equation}
and the spectral amplitude $A(\vett{q})$ is assumed to be a Gussian distribution, namely
\begin{equation}
A(\vett{q})=\frac{(\mathrm{det}\mathbf{V})^{1/4}}{\pi^{3/4}}\exp\left[-i\vett{q}\cdot\vett{r}_0-\frac{1}{2}\vett{q}\cdot\left(\mathbf{V}\vett{q} \right)\right],
\end{equation}
where we defined $\vett{q}\equiv\vett{k}-\vett{k}_0$ and $\vett{r}_0$ accounts for some possible initial transversal beam displacement. The matrix $\mathbf{V}$ is given, in the frame $(Ox_iy_iz_i)$, by
\begin{align}
\mathbf{V}=& \left(
\begin{array}{ccc}
 1/\sigma_1& 0 & 0 \\
0 & 1/\sigma_2 & 0 \\
0 & 0 & 1/\sigma_3 
\end{array}
\right),
\end{align}
being $\sigma_{1,2}$ the width of the Gaussian wave packet in the transversal plane $\{k_{x_i},k_{y_i}\}$ and $\sigma_3$ the width of the Gaussian wave packet in the longitudinal direction $k_{z_i}$. 

By making the change of variables $\vett{q}=\vett{k}-\vett{k}_0$ in Eqs. \eqref{parti} and by expanding in a Taylor series all the terms that appear in the integrals as a function of $\vett{k}_0+\vett{q}$ (namely the reflection coefficients $r_{\mu}(\vett{k}_0+\vett{q})$ and the polarization functions $\alpha_{\mu}(\vett{k}_0+\vett{q})$) up to the first order for the spatial and the Berry shift and up to second order for the $z$-dependent part, and by representing the (generally complex-valued) reflection coefficients using the polar representation of a complex number as
\begin{equation}
r_{\mu}(\vett{k})=|r_{\mu}(\vett{k})|e^{i\phi_{\mu}(\vett{k})}\equiv R_{\mu}(\vett{k})e^{i\phi_{\mu}(\vett{k})},
\end{equation}
we obtain the following results:
\begin{subequations}\label{result1}
\begin{align}
\langle\vett{R}\rangle_S=& \; \sum_{\mu=p,s}\left(\uvett{x}\, w_{\mu}\frac{\partial\phi_{\mu}}{\partial k_x}-\uvett{y}\, w_{\mu}\frac{\partial\phi_{\mu}}{\partial k_y} \right) - \uvett{y}\,\frac{w_pa_s^2 + w_sa_p^2}{a_pa_s} \cot\theta\sin\phi, \\
\langle\vett{R}\rangle_B=& \; - \, \uvett{y}\left(w_p w_s \right)^{1/2}\cot\theta\sin(\phi-\phi_p-\phi_s),
\\
\langle\vett{R}\rangle_A=& \; -\sum_{\mu=p,s}\left( \uvett{x} \, S_1^2w_{\mu}\frac{\partial\ln R_{\mu}}{\partial k_x}-\uvett{y} \, S_2^2 w_{\mu}\frac{\partial\ln R_{\mu}}{\partial k_y} \right)+\uvett{y}S_2^2 \frac{w_pa_s^2 -  w_sa_p^2}{a_pa_s}\cot\theta\cos\phi,
\end{align}
\end{subequations}
where
 we have defined the weighting factors $w_{\mu}$ (namely the fraction of energy contained in each polarization) as
\begin{equation}
w_{\mu}=\frac{a_{\mu}^2R_{\mu}^2}{a_p^2R_p^2+a_s^2R_s^2},
\end{equation}
and the factors $S_i^2$ ($i=\{1,2,3\}$) are defined in the following manner: 
\begin{equation}
S_i=\frac{\sigma_i}{k_0}\equiv\frac{\theta_i}{\sqrt{2}},
\end{equation}
where $\theta_i$ is the angular divergence of the wave packet along the direction $x_i$ (with $x_1\equiv\uvett{x}$, $x_2\equiv\uvett{y}$ and $x_3\equiv\uvett{z}$).

Summing all these contributions we obtain the final expression for the center of mass of the energy density distribution, that can be written as
\begin{equation}
\langle\vett{R}\rangle=\sum_{\mu=p,s}\left[\uvett{x} \, \Delta^\text{GH}_{\mu}(z)+\uvett{y} \, \Delta^\text{IF}_{\mu}(z)\right],
\end{equation}
where the \GH and Imbert-Fedorov shifts are given by the following formulas:
\begin{subequations}
\begin{align}
\Delta_{\mu}^\text{GH}(z)=& \, w_{\mu}\left(\frac{\partial\phi_{\mu}}{\partial k_x}-z \, S_1^2\frac{\partial\ln R_{\mu}}{\partial k_x}\right)\\
\Delta^\text{IF}_{\mu}(z)= & \, -w_{\mu}\left(\frac{\partial\phi_{\mu}}{\partial k_y}-z \, S_2^2\frac{\partial\ln R_{\mu}}{\partial k_y}\right)-\bigg[\frac{w_pa_s^2 + w_sa_p^2}{a_pa_s} \sin \phi - z \, \frac{w_pa_s^2 - w_sa_p^2}{a_pa_s}\cos\phi   \nonumber \\
& \;-  \left(w_p w_s \right)^{1/2}\sin(\phi-\phi_p-\phi_s)\bigg]\cot\theta.
\end{align}
\end{subequations}
Note that in writing these formulas we retained the terms depending on the derivative with respect to $k_y$, often neglected because they are  zero for Fresnel coefficients of a planar interface. Here, however, we kept them for the sake of completeness.

Note, moreover, that the results presented in this section give the same expression of the GH and IF shift of a monochromatic beam of light, despite they refer to the non monochromatic case. This fact is due to the fact that we stopped our analysis to the first order (in Taylor expansion) for the spatial and the Berry term and second order for the angular term, while, as stated in Ref. \cite{aielloW}, non monochromatic corrections to the GH and IF shifts should manifest (as an explicit dependence of the formulas above on the longitudinal width $S_3$) starting from the fourth order in the expansion with respect to $\vett{k}-\vett{k}_0$.

\section{Conclusions}
In conclusion, in this work we extended the concepts of \GH and Imbert-Fedorov shifts to the case of a non-monochromatic beam of light, namely a bounded wave packet. We shown that the expression of the center of mass of the energy density distribution can be easily written in a form similar to the monochromatic case (where the presence of the weighting factor $k/|k_z|$ is the signature of non monochromaticity) and we applied these results to the simple case of a quasi monochromatic Gaussian beam, showing that the non monochromatic nature of the beam does not play a role up to the first and second order terms of a suitable perturbative expansion, for the spatial and the angular part, respectively. These results may appear simple at a first glance, but in reality they are not, since conceptually it is not obvious that at the lowest order the non monochromatic nature of a field does not play any role in determining its shifts upon reflection. Moreover, since the expression of the center of mass for a non monochromatic wave packet of light (Eq. \eqref{centro}) is sensibly different from its monochromatic counterpart, it is not trivial at all to expect that these modification do not count or exactly compensate. Moreover, the presence of the weighting factor $k/|k_z|$ makes the integrals not solvable anymore analytically if this factor is not expanded in series. This is a signature of the perturbative action of the non monochromaticity on the process of reflection, and it clearly states that the major contribution comes from the zeroth order, i.e. from the monochromatic results. In future works, however, we intend to study deeper this problem, trying to understand under which conditions (and at what order) the non monochromatic corrections start to play a role.


\begin{thebibliography}{}
\bibitem{nota} The same reasoning is valid for the term containing $y\campo{E}$.
\end{thebibliography}
\end{document}